\documentclass{mem}
\usepackage{natbib}\usepackage{txfonts}\usepackage{balance}
\usepackage{graphicx,bm}
\usepackage[a4paper,breaklinks,dvipdfm]{hyperref}
\idline{75}{282}
\begin{document}
\def\teff{$T\rm_{eff }$}
\def\kms{$\mathrm {km s}^{-1}$}
\newcommand{\vect}[1]{\boldsymbol{#1}}
\newcommand{\fquad}[2]{#1^T #2 \: #1}
\newcommand{\de}{\mathrm{d}}
\newcommand{\defin}{=}
\def\ol#1{\overline{#1}}
\def\db#1{\dot{\mathbf{#1}}}
\def\ddb#1{\ddot{\mathbf{#1}}}
\def\hb#1{\hat{\mathbf{#1}}}
\def\hbb#1{\hat{\bm{#1}}}
\def\vb#1{\vec{\mathbf{#1}}}
\def\vbb#1{\vec{\bm{#1}}}
\def\tt#1{\texttt{#1}}
\def\bb{\bibitem}
\def\ber{\begin{eqnarray}}
\def\eer{\end{eqnarray}}
\def\beq{\begin{equation}}
\def\eeq{\end{equation}}
\newcommand{\contrib}[3]{#1\quad\texttt{<#2>}{\small\\\quad\textit{#3}}\\[1ex]}
\def\rmd{{\rm d}}

\title{
Using Ring Laser Systems to Measure Gravitomagnetic Effects on Earth
}

   \subtitle{}

\author{
Matteo Luca \,Ruggiero\inst{1,2} 
          }

  \offprints{Matteo Luca Ruggiero}

\institute{
DISAT,  Politecnico di Torino,
Corso Duca degli Abruzzi 24,
10129 Torino, Italy
\and
INFN, Sezione di Torino
\email{matteo.ruggiero@polito.it}
}

\authorrunning{Ruggiero}

\titlerunning{Ring Laser ang Gravitomagnetic Effects}

\abstract{Gravitomagnetic effects originates from the rotation of the source of the gravitational field and from the rotational features of the observers' frame. In recent years, gravitomagnetism has been tested by means of its impact on the
precession of LAGEOS orbits and on the precession of spherical gyroscopes in the GP-B experiment. What we suggest here is that light can be used as a probe to test gravitomagnetic effects in an terrestrial laboratory:  the proposed detector consists of large ring-lasers arranged along three orthogonal axes.
}
\maketitle{}


Any  theory that combines Newtonian gravity with Lorentz invariance, such as general relativity does,  must include a gravitomagnetic (GM) field \citep{gem}: in particular, gravitomagnetism is related to the presence of mass currents in the reference frame of a given observer. Actually, it can be showed \citep{bini92,bini94} that  gravitoelectromagnetic (GEM) fields can be introduced whenever one applies splitting techniques: the field equations of general relativity  and geodesics equation can be recast in a 3+1 space+time form, in which they are  analogous to Maxwell's equations and Lorence force law. In other words, when the family of observers is defined, it is possible to obtain the corresponding GEM fields and use them to define the evolution of the physical quantities that can be measured by these observers. In particular,  the GM field is related to the off diagonal components $g_{0i}$ (gravitomagnetic components) of the metric tensor in the frame adapted to the family of observers; generally speaking, the GM field originates from the rotation of the source of the gravitational field and from the rotational features of the observers' frame. So, to properly speak of a the gravitomagnetic field, it is necessry to assume a  particular family of observers in a given space-time, and, consequently, the measurements of the gravitomagnetic field are referred to these observers. \\ For instance, in the
case of celestial bodies, including the Earth, and excluding translational
motion with respect to the center of the body, gravitomagnetic effects are
due to the absolute rotation of the massive source with respect to distant
stars. When the Einstein equations in vacuum are applied to this kind of
symmetry and are linearized (weak-field approximation) GM is accounted for
by the analogue of a magnetic field of a rotating spherical charge. 

The rotation of the source of the gravitational field affects a gyroscope orbiting around it, in such a way that it undergoes the so-called Lense-Thirring precession, or dragging of the inertial frames of which the gyroscope define an axis \citep{schiff,ciufoliniwheeler}. The Lense-Thirring effect was first measured by  Ciufolini \citep{LARES,LARES1}, who deduced the relativistic
precession of the whole orbital momentum of two LAGEOS satellites. The recent GP-B experiment was based
on the measurement of the precession of four freely falling
spherical gyroscopes, carried by a satellite in polar orbit around the Earth
\citep{GPB1}. A comprehensive review on the measurement of the Lense-Thirring effect can be found in \citet{iorio2011}.

A different experimental approach aimed at the detection of the gravitomagnetic effects consists in using light as a probe.  
In this case the main remark is that the propagation of light in the
gravitational field of a rotating body is not symmetric: the coordinated
time duration for a given space trajectory in the same sense as the rotation
of the central source is different from the one obtained when moving in the
opposite direction. This asymmetry would for instance be visible in the
Shapiro time delay of electromagnetic signals passing by the Sun (or
Jupiter) on opposite sides of the rotation axis of the star (or the planet)
\citep{TRT2000,TNR2005}. In general, propagation of light is not symmetric whenever gravitomagnetic terms $g_{0i}$ are present in the metric tensor in the frame adapted to the family of observers: this property of the propagation of light is the one which we wish to
exploit in our Earth-bound experiment, called G-GranSasso \citep{bosi2011},  using a set of ring lasers.  In a terrestrial laboratory light exiting a laser cavity in opposite directions
is forced, using mirrors, to move along a closed path in space, and it experiences the gravitational field in the vicinity of the world-line of the laboratory. The two  directions are
not equivalent since gravitomagnetic terms are present in the metric tensor describing the gravitational field in the laboratory and, consequently, an observer in the laboratory would notice that the two  proper-times  required for light to come back to the
active cavity are (slightly) different. When the appropriate background metric describing the gravitational field of the rotating Earth is defined (see \citet{bosi2011}), it is possible to show that the proper-time difference between the two propagation times turns out to be 
\begin{equation}
\delta \tau = \frac{4}{c^{2}}\mathbf{A} \cdot \mathbf{\Omega },  \label{eq:flusso}
\end{equation}
where   $\mathbf A=A\vect{u}_n$ is the area enclosed by the beams and oriented according to its normal vector $\vect{u}_n$.
In particular, $\bm \Omega$ is simply related to the gravitomagnetic field in the laboratory frame, and it is $\bm \Omega=\bm \Omega_{\oplus}+\bm \Omega'$;  the term proportional to $\bm \Omega_{\oplus}$ is the purely kinematic Sagnac term, due to the rotation of the Earth, while $\bm \Omega' = \bm \Omega_{G}+ \bm \Omega_{B}+  \bm \Omega_{W} +\bm\Omega_{T}$ encodes the relativistc effects
\begin{eqnarray}\bm \Omega_{G}&=&
-\left( 1+\gamma \right) \bm \nabla U(R) \wedge \bm V, \label{eq:OmegaDS} \\
\bm \Omega_{B}&=&-\frac{1+\gamma +\alpha_{1}/4}{2} \left(\frac{\bm J_{\oplus}}{R^{3}}-\frac{3 \bm
 J_{\oplus} \cdot \bm R}{R^{5}}\bm R \right),\label{eq:OmegaLT} \\
\bm \Omega_{W} & = & \alpha_{1}\frac{{1}}{4} \bm \nabla U(R) \wedge \bm W, \label{eq:Omegaw} \\
\bm \Omega_{T}&=&-\frac{1}{2} \bm V \wedge  \frac{d \bm V}{dT}. \label{eq:OmegaTh}
\end{eqnarray}
where $-U(R)$ is the Newtonian potential, $\bm J_{\oplus}$ is the angular momentum of the Earth,
$W_{i}$ is the velocity of the reference frame in which the Earth is at rest with respect to mean
rest-frame  of the Universe; $\gamma$ and $\alpha_{1}$ are post-Newtonian parameters that measure,
respectively, the effect of spatial curvature and the effect of preferred frames. All  terms in (\ref{eq:OmegaDS})-(\ref{eq:OmegaTh}) must be evaluated along the laboratory world-line (hence, they are constant in the local frame),
whose position and velocity in the background frame are $\bm R$ and $\bm V$, respectively.
In particular $\bm \Omega'$ is made of four contributions:
 i) the geodetic or de Sitter precession $\bm \Omega_{G}$; ii) the Lense-Thirring precession $\bm\Omega_{B}$; iii)  $\bm \Omega_{W}$ is due to the preferred frames effect;
and iv) the Thomas precession $\bm \Omega_{T}$.

For a ring laser in an Earth-bound laboratory, the geodetic and Lense-Thirrng terms are both of order $\sim 10^{-9}$
with respect to the Sagnac term, while the Thomas term is 3 orders of magnitude smaller.  As for the preferred frames term, the best estimates (see e.g. \citet{bell,damour96}) show that this effect is about 2 orders of magnitude smaller than the geodetic and Lense-Thirring terms. Consequently,  to leading order, the relativistic contribution to the rotation measured by the ring laser turns out to be $ \bm \Omega_{REL} = \bm \Omega_{G}+ \bm \Omega_{B}$. 

Since $\Omega_{REL} \simeq  10^{-9}\ \Omega _{\oplus }$, angles between vectors must be measured
at the corresponding accuracy level. Unfortunately, the absolute measurement
of $\boldsymbol{u}_{n}$ in the fixed stars reference system with the
accuracy of nano-radians can hardly be achieved. The core idea of the G-GranSasso experiment is to  relax this
requirement by using $M\geq3$ ring lasers oriented along directions $\boldsymbol{ u^{\alpha }}$
 ($\alpha =1\dots M$), where not all $\boldsymbol{u}^{\alpha }$
lie in the same plane. In fact, the vector $\boldsymbol{\Omega }$ can be
completely measured by means of its projections on at least 3 independent
directions.
In conclusions, we propose a terrestrial experiment  to detect the general relativistic effects due to the curvature of space-time around the Earth (de Sitter effect) and to rotation of the planet (dragging of the inertial frames or Lense-Thirring effect). It is ultimately based on the comparison between the IERS value of the Earth rotation vector and corresponding measurements obtained by a three-axial laser detector of rotation. In particular, the proposed detector consists of six large ring-lasers arranged along three orthogonal axes. With shot noise limited square rings of 6 m side, which can achieve a sensitivity of $20\ prad/s/\sqrt{Hz}$ and 2 years integration time, the $1\%$ sensitivity required for the measurement of the Lense-Thirring drag can be reached.


\bibliographystyle{aa}

\end{document}